{}%
{}%
\documentclass[english,aps,prd,superscriptaddress,preprintnumbers,nofootinbib,preprint]{revtex4-1} %
\usepackage{amsmath} %
\usepackage{amssymb} %
\usepackage{amsfonts} %
\usepackage{graphicx} %
\usepackage{xcolor}
\usepackage[labelformat=simple]{subcaption}

\usepackage[utf8]{inputenc}
\usepackage[english]{babel}

\usepackage{cleveref}
\crefformat{section}{Section~#2#1#3}
\crefformat{figure}{Fig.~#2#1#3}
\crefformat{table}{Table~#2#1#3}
\crefformat{equation}{Eq.~(#2#1#3)}
\crefrangeformat{equation}{Eqs.~(#3#1#4) to~(#5#2#6)}
\crefmultiformat{equation}{Eqs.~(#2#1#3)}{ and~(#2#1#3)}{, (#2#1#3)}{ and~(#2#1#3)}
\crefrangemultiformat{equation}{Eqs.~(#2#1#3)}{ and~(#2#1#3)}{, (#2#1#3)}{ and~(#2#1#3)}
\crefname{appsec}{Appendix}{Appendices}

\newcommand{\be}{\begin{equation}}
\newcommand{\ee}{\end{equation}}

\allowdisplaybreaks

\begin{document} 

\preprint{TTP20-007, TUM-HEP-1251/20, TIF-UNIMI-2020-5} %
\def\TTP{Institute for Theoretical Particle Physics, KIT, Karlsruhe, Germany} %
\def\Milan{Dipartimento di Fisica, Universit\`a di Milano and INFN, Sezione di Milano, Via Celoria 16, I-20133 Milano, Italy} %
\def\IKP{Institut f\"ur Kernphysik, KIT, 76344 Eggenstein-Leopoldshafen, Germany} %
\def\CERN{Theoretical Physics Department, CERN, 1211 Geneva 23, Switzerland} %
\def\TUM{Physik-Department T31, Technische Universität München, James-Franck-Straße 1, D-85748 Garching, Germany} %
\def\Padua{Dip. di Fisica e Astronomia ``Galileo Galilei'', University of Padova, Via Marzolo 8, I-35131, Padova, Italy} %
\def\PaduaINFN{INFN, Sezione di Padova, Via Marzolo 8, I-35131, Padova, Italy} %

\title{Two-Loop QCD-EW Master Integrals for Z Plus Jet Production at Large Transverse Momentum} %

\author{Hjalte Frellesvig} %
\email[Electronic address: ]{hjalte.frellesvig@pd.infn.it} %
\affiliation{\TTP}\affiliation{\Padua, \PaduaINFN} %

\author{Kirill Kudashkin} %
\email[Electronic address: ]{kirill.kudashkin@mi.infn.it} %
\affiliation{\TTP}\affiliation{\Milan} %

\author{Christopher Wever} %
\email[Electronic address: ]{christopher.wever@tum.de} %
\affiliation{\TUM} %

\begin{abstract} %
The production of electroweak $Z$ bosons that decay to neutrinos and recoil against jets with large transverse momentum $p_\perp$ is an important background process to searches for dark matter at the Large Hadron Collider (LHC). To fully benefit from opportunities offered by the future high-luminosity LHC, the theoretical description of the $pp \to Z+j$ process should be extended to include mixed QCD-electroweak corrections. The goal of this paper is to initiate the computation of such corrections starting with the calculation of the Feynman integrals needed to describe two-loop QCD-electroweak contributions to $q \bar q \to Z+g$ scattering amplitudes. Making use of the hierarchy between the large transverse momenta of the recoiling jet, relevant for heavy dark matter searches, and the $Z$ boson mass $m_{Z}$, we present the relevant master integrals as a series expansion in $m_{Z}/p_\perp$.
\end{abstract} %

\maketitle
\flushbottom

\section{Introduction} 
\label{sec:intro}

Studies of electroweak vector bosons  play an important role in
experiments at the Large Hadron Collider (LHC).
Large cross sections of processes with one and even two electroweak
gauge bosons
and their  clean leptonic decay signatures allow many precision tests of
the Standard Model (SM)~\cite{Aaboud:2017hbk,Sirunyan:2018cpw}. However, these
same large cross sections imply that processes with  vector bosons are
important backgrounds to searches for physics beyond the Standard Model
(BSM). For dark matter searches specifically,
the most important among them
is $pp \to Z(\to \nu \bar \nu) + j$ (with $j$ denoting a hadronic jet), where the produced neutrinos escape detection and lead to the missing
energy signature~\cite{Sirunyan:2017jix,Aaboud:2017phn}.
It was pointed out in Ref.~\cite{Malik:2013kba} that
understanding
this process to the degree required for dark matter searches at the LHC
can not rely on purely data-driven techniques and that
theoretical input is required. Interestingly, in spite of the
fact that theoretical studies of $Z+j$ production have a long
and  successful history, it was argued recently~\cite{Lindert:2017olm} that
one important ingredient is still missing.

Indeed, to search for  heavy ${\cal O}(1~{\rm TeV})$ dark matter
particles, we look for events with high-$p_\perp$ jets and large missing
energy;
for the main background process $pp \to Z+j$, this ${\cal O}(1~{\rm TeV})$
missing energy is comparable to the transverse momentum of the vector
boson $p_{\perp}$~\cite{Abercrombie:2015wmb}.  It is well-known that  for
processes  with large $p_\perp$,
electroweak corrections get enhanced by  electroweak Sudakov
logarithms $ \log^2 p_\perp/m_{Z}$.
Hence, although the QCD corrections play a much more important role at
small $p_\perp$ than the
electroweak ones, this hierarchy becomes less obvious at high transverse
momentum.
As a consequence, a recent state-of-the-art theoretical analysis of vector boson plus jet production~\cite{Lindert:2017olm}
came to the conclusion
that unknown mixed QCD-electroweak corrections are among the largest
sources of uncertainty in a theoretical description of $Z+j$ production
at high $p_\perp$ at the LHC. Developing a better understanding of
these corrections is the long-term physics goal of this paper.

The theoretical understanding of $Z+j$ production
in hadron collisions is very advanced. Indeed, the QCD corrections to  $Z+j$ production are known to an impressive next-to-next-to-leading order (NNLO)~\cite{Ridder:2016nkl,Boughezal:2016isb,Boughezal:2016dtm}; at high $p_\perp$, these corrections increase the next-to-leading-order (NLO) predictions
by about $50\%$. The electroweak corrections have been computed up to NLO~\cite{Denner:2009gj,Denner:2011vu,Kallweit:2015dum} and  electroweak Sudakov logarithms, relevant at high $p_\perp$ are known through next-to-leading logarithmic terms at NNLO in the weak coupling constant~\cite{Kuhn:2004em,Kuhn:2005az,Kuhn:2007qc}.

The relevance of mixed QCD-electroweak corrections for dark matter searches
was quantified in Ref.~\cite{Lindert:2017olm} by comparing a
multiplicative and additive prescription for combining
NLO QCD and NLO electroweak corrections. It was found that the
difference between the two prescriptions leads to a $5-10\%$ ambiguity in the total cross section at high $p_{\perp}$~\cite{Lindert:2017olm}. Since the NLO QCD $K$-factors are large and the EW corrections can amount to up to a few tens of percents due to the EW Sudakov logarithms~\cite{Kuhn:2004em,Kuhn:2007cv}, it is crucial to accurately compute the NNLO mixed QCD-EW corrections to achieve few-percent accuracy for future high-luminosity LHC (HL-LHC) measurements. These corrections involve unknown two-loop virtual Feynman integrals with massive propagators and an off-shell leg.

\begin{figure}[!t]
\centering
\includegraphics[width=0.3\linewidth]{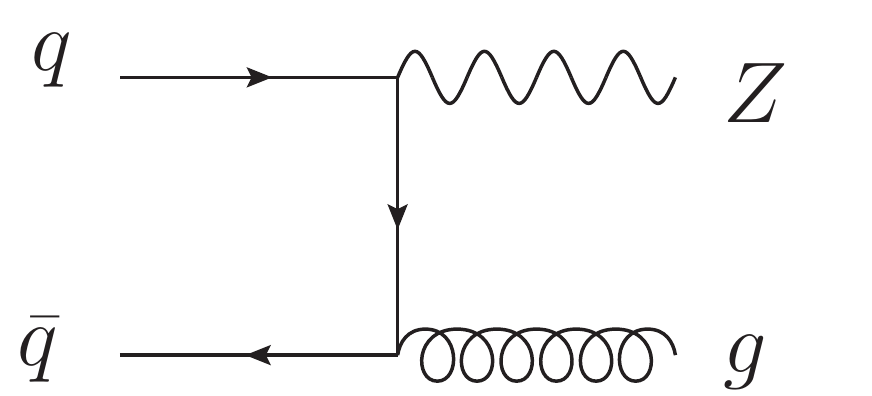} \;
\includegraphics[width=0.3\linewidth]{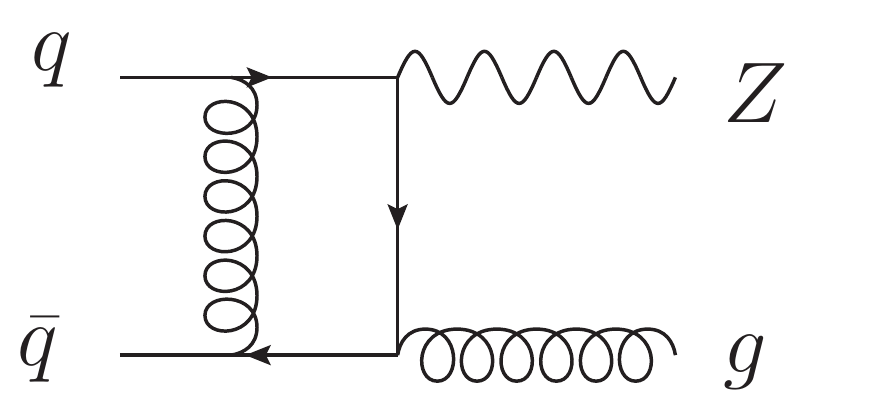} \;
\includegraphics[width=0.3\linewidth]{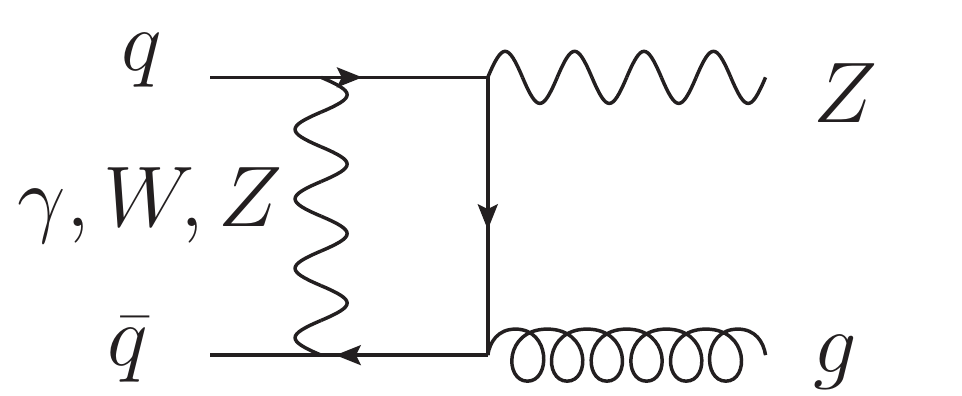} \\
\includegraphics[width=0.3\linewidth]{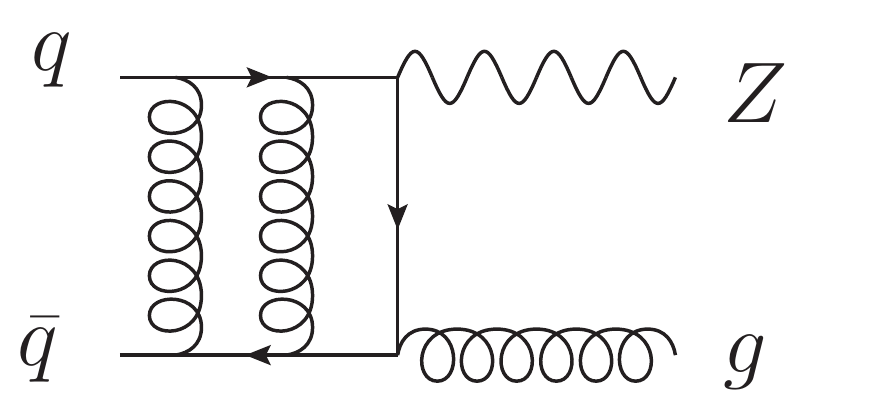} \;
\includegraphics[width=0.3\linewidth]{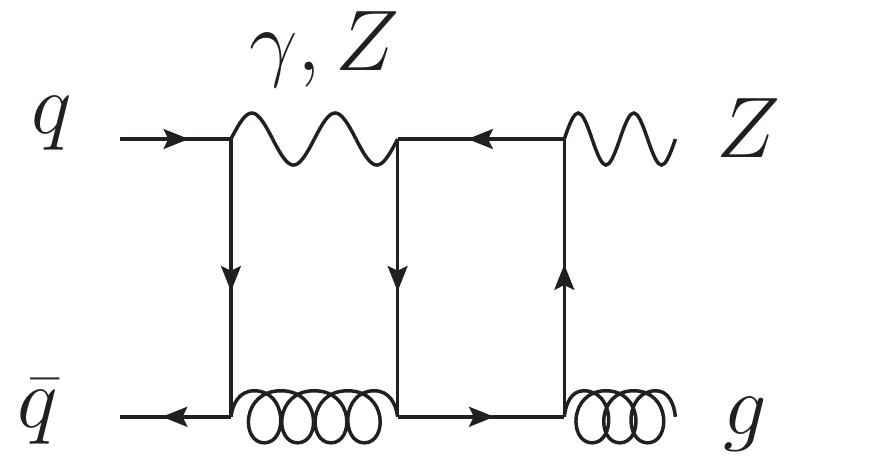} \;
\includegraphics[width=0.3\linewidth]{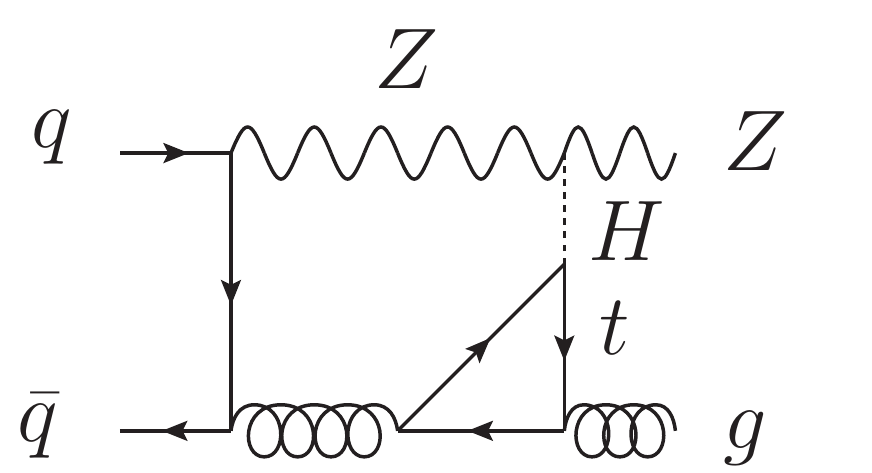} \\
\caption{Examples of Feynman diagrams that contribute to $Z+j$ production at LO (top-left) and virtual corrections at NLO QCD (top-middle), NLO EW (top-right), NNLO QCD (bottom-left) and mixed NNLO QCD-EW (bottom-middle and -right).}
\label{fig:order-diagrams}
\end{figure}

In Figure~\ref{fig:order-diagrams} we show some representative $2\rightarrow 2$ Feynman diagrams that contribute to $Z+j$ production at the LHC at various orders in QCD and EW. The dominant partonic contributions at LO are the $q\bar{q}\rightarrow Zg$ and $qg\rightarrow Zq$ processes. The majority of the missing two-loop mixed QCD-EW diagrams are those where the LO graphs are supplemented with an extra gluon, and either a massive $V=W^{\pm},Z$ or massless photon exchange. There are also contributions from closed quark loops (cf. bottom middle graph in Figure~\ref{fig:order-diagrams}), with either a massive (top) or massless quark-loop. 

In this paper we will include contributions of all massless quarks but systematically neglect the contributions that involve
the top quark, since these contributions contain another scale and it is very difficult to compute them. However, at (very)
large $p_\perp$, the top-quark contribution may, to some extent, be approximated by that of a massless quark and we can check to what
degree the top quark is relevant by simply including it in the massless
approximation. While not ideal, this will provide us an estimate of how
relevant the top-quark contribution is for mixed QCD-EW
corrections to $Z+j$ production
at high $p_\perp$.

In what follows, we compute the master integrals (MI) contributing to mixed QCD-EW corrections to $Z+j$, treating
all quarks as massless, in the limit of {\it large} $p_{\perp}$. Note that
in the massless quark approximation,
exchanges of a Higgs boson (cf. bottom-right graph in Figure~\ref{fig:order-diagrams}) do not contribute.
Furthermore, we take the $Z$ and $W$ boson masses to be equal,
$m_{W^{\pm}}=m_Z$. A rough estimate of the error introduced by such an approximation is $(m_Z^2-m_W^2)/m_Z^2\sim 20\%$ on the QCD-EW corrections, which corresponds to an error of order $1\%$ or less on the full $Z+j$ prediction.
The only place where this approximation
may  be questioned  is in the arguments of the Sudakov logarithms since
their mass difference is not $p_\perp$-suppressed. However,
even in this case, it leads to a tiny relative modification of
QCD-EW corrections by  $(m_Z-m_{W^{\pm}})/(m_Z \log (p_\perp^2/m_Z^2)) \sim
{\cal O}(5 \%)$. Since this expansion improves at higher $p_{\perp}$, our results are certainly valid within $1\%$ in the relevant $p_{\perp}\sim 600-1500$ GeV range where the statistical error is expected to be about $1-10\%$ at the HL-LHC and the theory errors associated to the missing QCD-EW corrections are about $2-5\%$~\cite{Lindert:2017olm}.

The method of differential equations (DE) has been very fruitful for computing MI~\cite{Kotikov:1990kg,Kotikov:1991pm,Bern:1992em,Remiddi:1997ny,Gehrmann:1999as,Argeri:2007up,Henn:2013pwa}. We will follow Ref.~\cite{Kudashkin:2017skd}, where an algorithmic method for expanding in $m_{\text{top}}^2/(p_{T}^H)^2\ll 1$ was used to compute the two-loop virtual amplitudes to the $H+$jet process at large Higgs transverse momentum $p_{T}^H$ including finite top-mass $m_{\text{top}}$ effects (see also~\cite{Mueller:2015lrx,Melnikov:2016qoc,Melnikov:2017pgf,Davies:2018qvx,Davies:2018ood} for similar expansions). In this method, the DE satisfied by the MI are expanded directly in the small parameter $m_Z^2/p_{\perp}^2$. This method has consistently shown to reduce the complexities of computing MI with massive propagators~\cite{Bonciani:2019jyb}. The MI presented in this paper are expanded through next-to-leading-power in the small parameter $m_Z^2/p^2_{\perp} \sim 10^{-2}$ for $p_{\perp} \sim 1~{\rm TeV}$. 

The remainder of the paper is organized as follows. In Section~\ref{sec:topologies}, we introduce the notation used in this paper and describe the integral families required for computing the mixed QCD-EW virtual $Z+j$ amplitude (with all quarks massless). Section~\ref{sec:masters} summarizes the method of using differential equations to compute the MI as an expansion in $m_Z^2/\hat{s}_{ij}$, while the calculation of the boundary constants to fix the MI is explained in Section~\ref{sec:bound}.  In Section~\ref{sec:crossings} we give some brief details on how to analytically continue and cross our MI solutions to other production channels and in Section~\ref{sec:checks} we report on various numerical checks that were performed for the MI solutions. Finally, we conclude in Section~\ref{sec:conc}. Alongside this paper, we include ancillary files that contain the solutions for our MI.

\section{Definitions and Topologies} 
\label{sec:topologies}

As discussed in the introduction, we consider Feynman integrals that are needed to describe the mixed QCD-EW two-loop corrections to $Z+j$ production, in the approximation when all quarks are massless. We also take the vector boson masses to be equal, $m_V := m_Z = m_W$ and consider the external vector boson to be on the mass shell.

\subsection{Kinematics}

We choose the kinematics such that the three partons have (incoming) momenta $p_1$, $p_2$, $p_3$, and then the kinematic quantities are defined as
\begin{align}
p_1^2 = p_2^2 = p_3^2 = 0\,, \quad (p_1{+}p_2{+}p_3)^2 = m_V^2\,, \nonumber \\[-11mm] \nonumber
\end{align}
\begin{align}
(p_1{+}p_2)^2 = s \,, \quad (p_1{+}p_3)^2 = t \,, \quad (p_2{+}p_3)^2 = u = m_V^2 {-} s {-} t \,.
\end{align}
Thus there are three scales in our problem: $s$, $t$, and $m_V$. From these we can define an overall dimensionful scale, and two dimensionless quantities
\begin{align}
\sigma \equiv -s \,, \quad \chi \equiv \frac{t}{s} \,, \quad \mu \equiv \frac{m_V^2}{-s} \,.
\end{align}

In the following we will mostly consider $u$-channel kinematics (corresponding to the particles with momenta $p_2$ and $p_3$ being incoming). This implies
\begin{align}
s,t < 0, \quad u, m_V^2>0
\end{align}
or correspondingly
\begin{align}
\sigma,\chi,\mu > 0.
\end{align}

In the $u$-channel we have
\begin{align}
|p_T| = \sqrt{ts/u \,} = \sqrt{\sigma \chi/( 1 + \chi + \mu)}
\end{align}
and high transverse momentum corresponds therefore to $\chi \sim \mathcal{O}(1)$ and $|\mu| \ll \{1,|\chi|\}$, justifying the expansion we use in our computations.

\subsection{Integral families}

\begin{figure}[!ht]
\centering
\includegraphics[width=1.6cm]{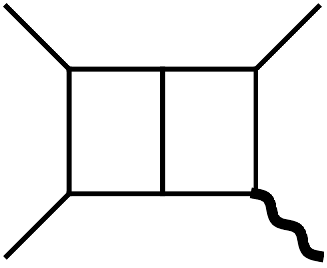} \;
\includegraphics[width=1.6cm]{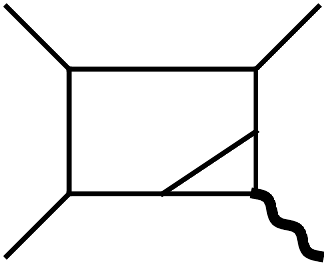} \;
\includegraphics[width=1.6cm]{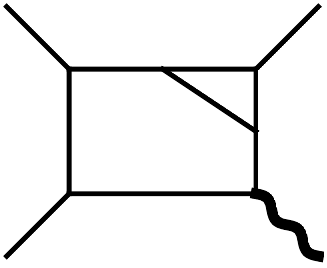} \;
\includegraphics[width=1.6cm]{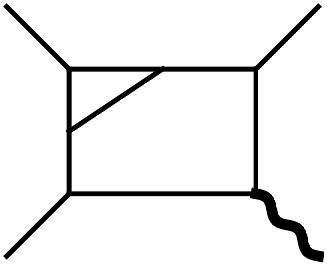} \;
\includegraphics[width=1.6cm]{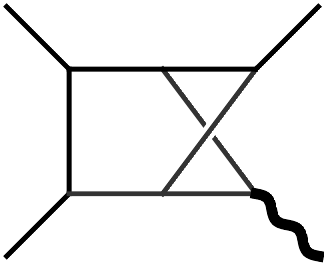} \;
\includegraphics[width=1.6cm]{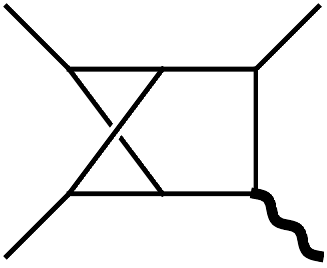} \\
\caption{The different top-sector Feynman integrals, in terms of the distribution of the external massive leg.}
\label{fig:externalleg}
\end{figure}

In order to find the classes of integrals that need to be evaluated, we first look at the positioning of the external $Z$ boson. We find that there are six options (up to crossings) for genuine seven-propagator sectors, shown in Fig. \ref{fig:externalleg}. These cases can be fitted into two 9-propagator families, a planar and a non-planar. We choose the momenta of the propagators in these families as
\begin{align}
q_1^{\text{pl}} &= k_1, & q_2^{\text{pl}} &= k_1 {+} p_1, & q_3^{\text{pl}} &= k_1 {+} p_1 {+} p_2, \nonumber \\
q_4^{\text{pl}} &= k_2 {+} p_1 {+} p_2, & q_5^{\text{pl}} &= k_2 {-} p_3, & q_6^{\text{pl}} &= k_2, \label{eq:familyp} \\
q_7^{\text{pl}} &= k_1 {-} k_2, & q_8^{\text{pl}} &= k_1 {-} p_3, & q_9^{\text{pl}} &= k_2 {+} p_1, & \nonumber
\end{align}
for the planar sectors and
\begin{align}
q_1^{\text{np}} &= k_1, & q_2^{\text{np}} &= k_1 {+} p_1, & q_3^{\text{np}} &= k_1 {+} p_1 {+} p_2, \nonumber \\
q_4^{\text{np}} &= k_2 {+} p_1 {+} p_2, & q_5^{\text{np}} &= k_2 {-} p_3, & q_6^{\text{np}} &= k_1 {-} k_2 {+} p_3, \label{eq:familynp} \\
q_7^{\text{np}} &= k_1 {-} k_2, & q_8^{\text{np}} &= k_2, & q_9^{\text{np}} &= k_1 {-} k_2 {-} p_2, \nonumber
\end{align}
for the non-planar sectors.

We note that this discussion has not regarded internal masses. For a mixed QCD-EW contribution, there has to be at least one internal electroweak boson in any diagram. Either there can be one internal electroweak boson connecting two quark-lines directly, or there can be two vector bosons that couple to the external vector boson and to quark-lines through the $W_+ W_- Z$ vertex.

\begin{figure}[!ht]
\centering
\includegraphics[width=3.5cm]{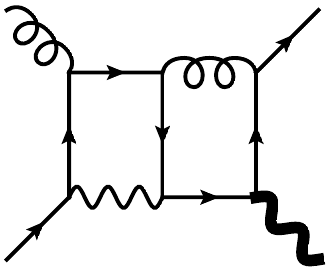} \;\;\;
\includegraphics[width=3.5cm]{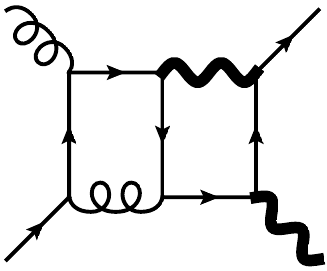} \;\;\;
\includegraphics[width=3.5cm]{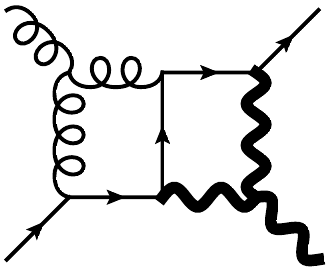} \\
\caption{Representative planar Feynman diagrams, contributing to the $0$-mass, $1$-mass, and $2$-mass cases respectively. Thick lines indicate massive propagators.}
\label{fig:massdistribution}
\end{figure}

Since the internal vector boson can be a photon, it is possible for a diagram to have no internal masses. For cases with one internal massive vector boson, it can be anywhere in the diagram except on a line connecting to the external $Z$, and since the momenta can always be chosen in such a way that the external $Z$ connects to the corner formed by momenta $q_4$ and $q_5$ (both in the planar and non-planar case) the single mass cases can have the mass in any propagator except in propagators labeled as 4 and 5 in Eqs.~(\ref{eq:familyp}, \ref{eq:familynp}). Finally the two-mass cases have to have the masses on propagators 4 and 5 simultaneously, see Fig.~\ref{fig:massdistribution}. This means that we have to consider 18 integral families in total: planar and non-planar families Eqs.~(\ref{eq:familyp}, \ref{eq:familynp}) modified by the presence of the internal mass $m_V^2$ on one of the following sets of propagators
\begin{align}
\Big\{ \{\},\, \{1\},\, \{2\},\, \{3\},\, \{6\},\, \{7\},\, \{8\},\, \{9\},\, \{4,5\} \Big\}.
\end{align}
With this in mind, our convention for the Feynman integrals is the following dimensionless combination
\begin{align}
I^f_{a_1,\ldots,a_9}(d,\chi,\mu) \; \equiv \; e^{2 \epsilon \gamma_E}\, \sigma^{a - d} \int \frac{d^d k_1 d^d k_2}{(i \pi^{d/2})^2} \, \frac{1}{P_{f1}^{a_1} \cdots P_{f9}^{a_9}},
\label{eq:idef}
\end{align}
where $\gamma_E$ is the Euler Mascheroni constant, $\epsilon = (4-d)/2$ is the dimensional regularization parameter, $a \equiv \sum_{i=1}^9 a_i$, and $P_{fi}$ is the $i$th propagator in a family $f$.
Our convention for the propagators is such that
\begin{align}
P_i = m^2 - q_i^2, \label{eq:prop}
\end{align}
where $m^2$ may be either $0$ or $m_V^2$ depending on the mass of the virtual particle.

\subsection{Integral reductions and master integrals}

It is well known that families of Feynman integrals can be expressed through a minimal basis of independent integrals, referred to as ``master integrals''. We perform the reductions to master integrals for each family, using IBP technology~\cite{Tkachov:1981wb, Chetyrkin:1981qh}, with the program Kira~\cite{Maierhoefer:2017hyi}. Since Kira is able to identify identical integrals in different families, we treat our problem as one big coupled system of linear equations rather than treating the families individually. In total we use a basis of $468$ master integrals, constructed in such a way that the integrals in the $7$-propagator sectors are independent under any permutation of $s$, $t$, and $u$. 

\section{Master Integrals from the differential equations} 
\label{sec:masters}

Integration by parts identities can be used to construct differential equations for the master integrals. To solve these equations in the kinematic regime $|\mu| \ll \{1,|\chi|\}$, we use a series ansatz~\cite{Beneke:1997zp}. The ansatz for an integral $f_n$ is of the form
\begin{align}
f_n(d,\chi,\mu) &= \sum_{ijk} c_{n,i,j,k} (d,\chi) \mu^{i+j \epsilon} \log^k (\mu).
\label{eq:muansatz}
\end{align}
We call each individual expansion term that appears in the ansatz in Eq.~\eqref{eq:muansatz} a {\it branch}. The accuracy of this approximate expression is formally determined by the maximal value $i_{\text{max}}$ included in the $i$-sum. Once $i_{\text{max}}$ is chosen, all other limits in the sum are fixed from the differential equations discussed below. For most of the integrals, the $i$-sum starts from $i=0$ but for some, divergences are present which make  the $i$-sum start from a value as low as $i=-3$. For the $j$ sum only negative values appear, with all integrals having values in the interval from $-4$ to $0$. For the log-terms only $k=0$ and $k=1$ appear. Finally for a few integrals $i$ and $j$ also appear with certain half-integer values~\cite{Kudashkin:2017skd}.

In summary for the most general case we have to consider the following values of $i,j,k$
\begin{align}
i \in \{-3, -\tfrac{5}{2}, \ldots, 0, \ldots, i_{\text{max}} \} \,, \quad j \in \{-4, -3, - \tfrac{5}{2}, \ldots, 0\} \,, \quad k \in \{0,1\} \,.
\end{align}
For the results provided with this paper, we choose $i_{\text{max}}=1$ for the top-sector integrals, which requires\footnote{The reason for this is that mass-suppressed coefficients of the lower sectors enter the DE of the top-sector at $i_{\text{max}}=1$. We therefore compute the $\mu$-series expansion of the lower sectors up to an $i_{\text{max}}$-value required for reaching $i_{\text{max}}=1$ for the top-sector integrals.} us to choose at least $i_{\text{max}}\geq 1$ for the lower sectors in order to achieve this level of accuracy.
 
We solve for the the $c$-coefficients of Eq.~\eqref{eq:muansatz} using the method of differential equations~\cite{Kotikov:1990kg}. We combine the $\nu=468$ MIs into a vector $\boldsymbol{f}$ and write
\begin{align}
\partial_{\mu} \boldsymbol{f} &= A(d, \chi, \mu) \boldsymbol{f}.
\label{eq:mudifeq}
\end{align}
Note that we can choose the order of the master integrals in such a way that the matrix $A$ takes a (lower) block triangular form, with the blocks consisting of integrals with the same set of propagators. Since all the entries of $A$ are rational functions of $\mu$, it follows from Eq.~\eqref{eq:muansatz} that terms with different $j$ or $k$ decouple; hence expanding the RHS of Eq.~\eqref{eq:mudifeq} in $\mu$ allows us to extract relations between the coefficients $c_{n,i,j,k}$ in Eq.~\eqref{eq:muansatz} for each block, $j$, and $k$ at a time. At the end we are left with only $\nu=468$ unfixed coefficients.

At this point we have a valid solution for the $\mu$ differential equation Eq.~\eqref{eq:mudifeq}. The next step is to insert that result into the corresponding $\chi$ differential equation for $\boldsymbol{f}$, which leads to a set of differential equations for the undetermined $c$-coefficients.
Introducing a new vector $\boldsymbol{g}$ for these unfixed coefficients, we get the differential equation
\begin{align}
\partial_{\chi} \boldsymbol{g} &= B(d, \chi) \boldsymbol{g}.
\label{eq:chidifeq}
\end{align}
As this is a one-scale system, it can easily be\footnote{Only for the integrals in the seven-propagator sectors, it was necessary to do this. For the lower sectors the equations were simple enough that it was possible to integrate up Eq.~\eqref{eq:chidifeq} with more traditional methods.} brought into a canonical form ~\cite{Henn:2013pwa} using automatic public tools such as Fuchsia~\cite{Gituliar:2017vzm}. We find
\begin{align}
\partial_{\chi} \tilde{\boldsymbol{g}} &= \epsilon \tilde{B}(\chi) \tilde{\boldsymbol{g}}
\label{eq:chidifeqcanonical}
\end{align}
along with rules for mapping between the old $(\boldsymbol{g})$ and new $(\tilde{\boldsymbol{g}})$ vectors. As the entries of $\tilde{B}$ are in a $d \log$-form, this equation system can easily be integrated, leading to results of the form 
\begin{align}
\tilde{g}_n = \sum_{w=0}^4 \epsilon^w \tilde{g}_n^{(w)}(\chi) \; + \; \mathcal{O}(\epsilon^5).
\end{align}
The solutions $g^{(w)}(\chi)$ are given in terms of generalized polylogarithms (GPLs)~\cite{Goncharov:1998kja} of $\chi$ with weights $w$ and entries $\bar{a}$, $G(\bar{a},\chi)$. Finally, to obtain a complete solution, boundary values for the integrals are needed. Their computation is discussed in Section~\ref{sec:bound}.

As mentioned in the previous Section, integrals can be separated into families with $0$ or $1$ internal masses, and integrals with $2$ internal masses adjacent to the external massive vector-boson.
For integrals in the first category, all polylogarithms can be expressed as GPLs with entries taken from\footnote{In our results as they appear in the ancillary files, we also have GPLs with the entries $-2$ and $-\tfrac{1}{2}$ appearing. However, integral transformations of the form discussed in ref.~\cite{Vollinga:2004sn} allow for the mapping of these GPLs to the HPL set, at the cost of introducing an argument different from $\chi$.} the list $\{0,1,-1\}$, making them equivalent to harmonic polylogarithms~\cite{Remiddi:1999ew}. We note that the entry $1$ is spurious, in the sense that our integrals have no divergence in the point $\chi=1$.
In the two-mass category, the entries are taken from $\{0,-1,r_{+},r_{-}\}$ where
\begin{align}
r_{\pm} = -\frac{1}{2} \pm \frac{\sqrt{3}}{2} i
\end{align}
are third roots of unity, and the roots of the cyclotomic polynomial $x^2 + x + 1$. Polylogarithms with such entries are discussed in Refs.~\cite{Ablinger:2011te, Henn:2015sem}.
In addition\footnote{This is in addition to the ``usual'' set of $i \pi$, $\log(2)$, $\zeta(3)$, and $\text{Li}_4(\tfrac{1}{2})$.} four special, transcendental constants appear in the result. They are
\begin{align}
K_2 &= \text{Im}(\text{Li}_2(\tfrac{1}{2} + \tfrac{\sqrt{3}}{2}i ))\,, & K_{3;1} &= \text{Im}(\text{Li}_3(\tfrac{1}{\sqrt{3}}i ))\,, \nonumber \\
K_{4;1} &= \text{Im}(\text{Li}_4(\tfrac{1}{2} + \tfrac{\sqrt{3}}{2}i ))\,, & K_{4;2} &= \text{Im}(\text{Li}_4(\tfrac{1}{\sqrt{3}}i ))\,.
\end{align}

The results have been checked numerically as discussed in Section~\ref{sec:checks}.

\subsection{The ancillary files}
\label{sec:anc}

Alongside the arXiv version of this paper we include two ancillary files. The first file \textit{definitions.txt} contains, in Mathemathica format, the definitions of the integral families, the master integrals, and a few other definitions that are relevant to interpret the results. 
The other file \textit{results.txt} contains the solutions for the master integrals. They are presented in the branch structure given by Eq.~\eqref{eq:muansatz}, and the branches are expanded such that each integral includes terms up to weight 4, as described above. Each branch contains an object ORD indicating the missing $\epsilon$-order, and branches that are not exact in $\mu$ contain an object ALARM indicating the first missing power in the $\mu$-expansion.

\section{Boundary Conditions} \label{sec:boundary_condition}
\label{sec:bound}

The differential equations in $\mu$ and $\chi$ in Eqs.~\eqref{eq:mudifeq},~\eqref{eq:chidifeq} fix the coefficients $c_{n,i,j,k}(d,\chi)$ that enter the $\mu$-expansion ansatz (cf. Eq.~\eqref{eq:muansatz}) up to $468$ integration ``constants'' that only depend on the dimensional parameter $\epsilon=(4-d)/2$. The differential equations relate the coefficients $c_{n,i,j,k}(d,\chi)$ in branches with different $i,n$-values but same $j,k$ to each other as explained in the previous Section. For this reason, the 468 constants that still need to be computed appear in branches of several master integrals and we may compute a suitable branch, in order to deduce the corresponding constant. We computed these constants either by matching the appropriate branches to massless $m_V=0$ master integrals, or using the so-called regularity conditions, or lastly by direct computation of specific branches at a convenient phase space-point $s, t, u$. We illustrate these three methods below.

\subsection{Massless solutions}

A subset of the constants are fixed by the requirement that the massless branch coefficient $c_{n,0,0,0}$ in the ansatz in Eq.~\eqref{eq:muansatz}, is equal to the massless master integral $f_n(m_V=0)$ that is obtained when the vector-boson mass $m_V$ that appears in the propagators (cf. Eq.~\eqref{eq:prop}) and in the off-shellness of the external leg, is set to zero at the integrand level. These massless double-box (both planar and non-planar) integrals have been previously computed in~\cite{Smirnov:1999wz,Henn:2013pwa,Tausk:1999vh,Anastasiou:2000mf,Argeri:2014qva} to weight 4 and we take the results for the coefficients $c_{n,0,0,0}$ from there.

\subsection{Regularity conditions}

For the planar master integrals, a large subset of the remaining constants is fixed by requiring that any spurious poles or branch-points that arise from solving the differential equations vanish. The physical poles and branch-points of Feynman integrals  can be understood from their cuts. Unlike the non-planar integrals, the planar Feynman integrals only have cuts in two of the three $s,t,u$ Mandelstam variables, cf. Fig.~\ref{fig:massdistribution}. As it happens, the differential equations for some of the planar master integrals allow solutions with poles and branch-points that do not correspond to their physical cuts. Requiring that these unphysical singularities vanish allows us to fix many constants in the planar sectors.

\subsection{Direct computation of integration constants}\label{sec:Direct_comp}

After using the massless branch and regularity conditions described above, we are still left with a few remaining constants, all of them appearing in branches of non-planar master integrals. We fix these remaining constants by computing suitable branches of some integrals where the required constants appear, either at a regular kinematic point or taking limiting values of $s,t$ corresponding to a physical pole of the corresponding Feynman diagrams. 
For the master integrals with up to {\it six} propagators, we chose to compute the relevant branches at a regular point, which was either $s=t=-1$ ($\chi=1$) or $s=-1,t=-2$ ($\chi=2$), both not corresponding to any physical pole of the master integrals. These branches were either computed by applying the method of expansion by regions~\cite{Beneke:1997zp,Smirnov:2002pj} to the Feynman-parametric representation of the master integrals, or by applying the Mellin-Barnes method. We refer to Ref.~\cite{Melnikov:2016qoc} for a detailed example of a computation of a specific branch of a seven-propagator integral through the use of Feynman parameters\footnote{We performed some of the parametric integrals with the Maple-based package HyperInt~\cite{Panzer:2014caa}.} and to Ref.~\cite{Kudashkin:2017skd} for an example of a computation of an integral with six propagators using the Mellin-Barnes method.

Unfortunately, the application of these methods did not allow us to compute ten constants needed for non-planar integrals with seven propagators. For these cases we computed the relevant branches in either the limit $-t\rightarrow 0_+$ or $u\rightarrow 0_+$. Since both limits correspond to physical poles of the non-planar master integrals, their relevant branches typically behave as $(-t)^{-1+l_1\epsilon}$ or $u^{-1+l_2\epsilon}$ in these two limits. The branches that we computed were {\it chosen in such a way} that the remaining unknown constants in the solutions multiplied a $\chi$-dependent factor that contained one (or both) of the two possible $t,u$ singularities and we chose to compute these branches in the limit where that $\chi$-dependent factor has a pole, either at $-t\rightarrow 0_+$ or $u\rightarrow 0_+$. %
We give now an example of a constant that we computed by considering the $-t\rightarrow 0_+$ limit of a relevant branch of a seven-propagator master integral, using the method of Mellin-Barnes.

\begin{figure}[t!]
\centering
\includegraphics[width=0.45 \linewidth]{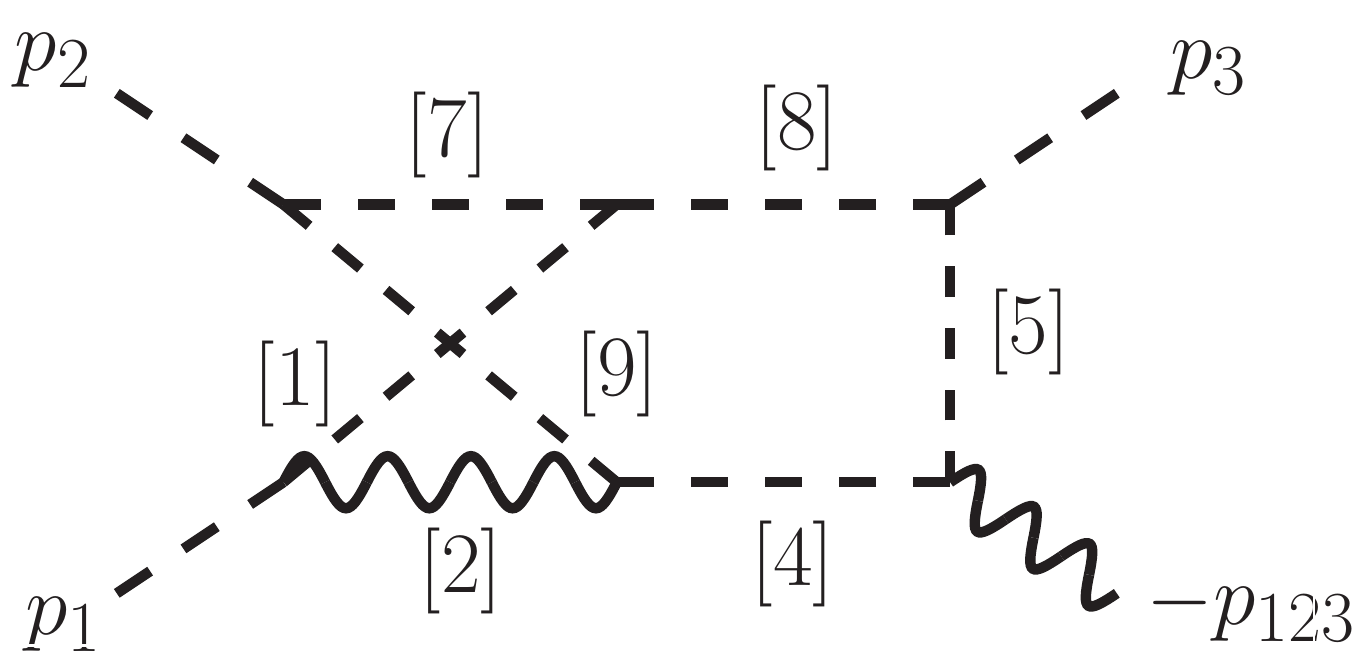}
\caption{The diagram corresponding to $I^{\rm np2}_{110110111}$. Dashed (solid) lines represent massless (massive) propagators and $[n]$ corresponds to the ordering in Eq.~\eqref{eq:familynp}.}
\label{fig::feyndiagNPL2t7}
\end{figure} 

Let us consider the master integral $I^{\rm np2}_{110110111}$, shown in Figure~\ref{fig::feyndiagNPL2t7},
\begin{gather}
I^{\rm np2}_{110110111} = \int 
\frac{\mathfrak{D}^dk \mathfrak{D}^dl}{k_1^2((k_1+p_1)^2-m_V^2)(k_2+p_{12})^2(k_2-p_3)^2(k_1-k_2)^2k_2^2(k_1-k_2-p_2)^2} \,, \label{eq:MB0}
\end{gather}
where the measure $\mathfrak{D}^dk\mathfrak{D}^dl$ is taken as in Eq.~\eqref{eq:idef}.
It has a branch $\mu^{-\epsilon}$, whose coefficient $c_{449,0,-1,0}$ contains a constant that multiplies a factor $\propto t^{-1}$. We are therefore interested in extracting that branch and computing its limit $-t\rightarrow 0_+$. The integral in Eq.~\eqref{eq:MB0} may be expressed through a Feynman-parametric representation with Symanzik polynomials as follows
\begin{eqnarray}
I^{\rm np2}_{110110111} &=& -\int_0^{\infty} \left(\prod_{i=1}^7 dx_i\right)\frac{\delta(1-x_1)\Gamma (2 \epsilon+3)}{\Gamma (\epsilon+1)^2} \mathcal{U}^{3 \epsilon+1} \mathcal{F}^{-2 \epsilon-3},\nonumber
\end{eqnarray}
with $u=m_V^2-s-t$ and $\kappa:=-1-i0$.\footnote{We introduce $\kappa$ in order to keep track of the proper analytic continuation when performing the Mellin-Barnes splittings introduced in Eq.~\eqref{eq:MB1} below.} The above Symanzik polynomials equal
\begin{eqnarray}
\mathcal{U}&=&(x_1+x_2) (x_3+x_4+x_5+x_6+x_7) +(x_3+x_4+x_6) (x_5+x_7), \\
\mathcal{F}&=&-i0+(-s) \big(x_1 x_3 (x_5+x_6)+x_6 (x_2 (x_3+x_7)+x_3 (x_5+x_7)) \big)+(-t) x_2 x_4 x_5+(-u) x_1 x_4 x_7 \nonumber\\
&& +m_V^2 \big(x_2 ((x_3+x_4+x_6) (x_5+x_7)+(x_1+x_2) (x_3+x_4+x_5+x_6+x_7))  \nonumber\\
& & +\kappa  x_4 (x_1 x_3+(x_5+x_7) x_3+x_2 (x_3+x_7)) \big).
\end{eqnarray}
We perform the integrals over the Feynman parameters by introducing Mellin-Barnes integrals, that split up terms inside the Symanzik polynomials as follows,
\begin{gather}
\frac{1}{(x+y)^\lambda}=\frac{1}{2\pi i}\int \limits_{-i\infty}^{+i\infty} dz \frac{y^z}{x^{z+\lambda}}\frac{\Gamma(-z)\Gamma(\lambda+z)}{\Gamma(\lambda)}.
\label{eq:MB1}
\end{gather}
The contour runs parallel to the imaginary axis in the complex $z$-plane and is chosen such that the singularities of $\Gamma(-z)$ and $\Gamma(\lambda+z)$ are  to the right (left), respectively of the integration contour. 
We need to introduce seven Mellin-Barnes integrals to be able to perform the integration over all seven Feynman parameters and we are left with,
\begin{gather} 
I^{\rm np2}_{110110111}=-\frac{\Gamma(-\epsilon)}{\Gamma (\epsilon+1)^2} \int \limits_{-i\infty}^{+i\infty} \left(\prod_{i=1}^7 dz_i\right)(-s)^{z_3} (-s-t)^{z_2} (-t)^{z_1} \kappa^{z_2+z_4}(m_V^2)^{-2 \epsilon-z_1-z_2-z_3-3} \nonumber\\
\times\, \frac{\Gamma (-z_1) \Gamma (-z_2) \Gamma (-z_4) \Gamma (-z_5) \Gamma (z_5+1) \Gamma (-z_6) \Gamma (-z_7) \Gamma (z_3-z_5+1) \Gamma (z_5-z_3)}{\Gamma (-z_4-z_6) \Gamma (-2 \epsilon+z_5+z_6) \Gamma (z_1+z_2+z_3+z_4+3) \Gamma (-\epsilon+z_1+z_2+z_3+z_4+2)} \nonumber\\
\times\, \Gamma (\epsilon+z_1-z_6+1) \Gamma (-\epsilon+z_5+z_6) \Gamma (z_1+z_2+z_4+1) \Gamma (z_2+z_5+z_7+1) \Gamma (-z_4-z_6+z_7) \nonumber\\
\times\, \Gamma (-2 \epsilon-z_2+z_6-z_7-1) \Gamma (z_2+z_3+z_4+z_6+2) \Gamma (-z_3-z_4-z_6+z_7-1) \nonumber\\
\times\, \frac{\Gamma (2 \epsilon+z_1+z_2+z_3+z_4+3)}{\Gamma (-z_3-z_4+z_5-z_6+z_7)}.
\label{eq:MB2}
\end{gather}
We put $-s=1$ and first extract the $\mu^{-\epsilon}=(m_V^2)^{-\epsilon}$ branch by closing appropriate integration contours to the left, picking up residues at $-2 \epsilon-z_1-z_2-z_3-3=-\epsilon$. Afterwards, we close contours such that we pick up residues at $z_1=-1+a\epsilon$ for any real-valued $a$. The latter residues correspond to the leading power pole\footnote{One finds by closing contours that there is no $t^{-2+a\epsilon}$ or higher pole.} at $-t\rightarrow 0_+$. Finally, we expand the result in $\epsilon$. All of these steps can be performed with the packages collectively known as MBTools \cite{MBTools}. After expanding the result in $\epsilon$, we are left with a sum of one-fold Mellin-Barnes integrals that may be performed by closing the contours either to the left or right, picking up a ladder of residues by the virtue of Cauchy's theorem. The resulting sum over residues may be then performed with the package XSummer~\cite{Moch:2005uc}. We note that, had we not taken the limit $-t\rightarrow 0_+$ but instead just evaluated the branch at a regular point, e.g. $-t=1$, the final expression after expanding in $\epsilon$ would have contained various three-fold Mellin-Barnes integrals that would have made the final calculation much more complicated.

The final result for the  $\mu^{-\epsilon}$ branch 
of the  $I^{\rm np2}_{110110111}$ at $s=-1$, in the limit $\chi=-t\rightarrow 0_+$ equals to weight four
\begin{gather}
\label{eq:MB3}
 I^{\rm np2}_{110110111} \supset  
\frac{\mu^{-\epsilon}}{\chi} \left\{-\frac{4}{\epsilon^4}+\frac{3 \log (\chi)-\frac{2}{3}-9 i \pi }{\epsilon^3}+\frac{-\frac{1}{2} \log ^2(\chi)+\left(\frac{2}{3}+6 i \pi \right) \log (\chi)+\frac{47 \pi ^2}{6}-\frac{25 i \pi }{3}+2}{\epsilon^2} \right.\nonumber\\
\left. +\frac{-\frac{1}{2} \log ^3(\chi)+\left(-\frac{1}{3}-i \pi \right) \log ^2(\chi)+\left(-2+\frac{25 i \pi }{3}-\frac{47 \pi ^2}{6}\right) \log (\chi)+22 \zeta_3+\frac{26 i \pi ^3}{3}+\frac{9 \pi ^2}{2}+23 i \pi -6}{\epsilon} \right. \nonumber\\
\left. +\frac{11 \log ^4(\chi)}{24}+\frac{1}{9} (1-6 i \pi ) \log ^3(\chi)+\left(1-\frac{25 i \pi }{6}+\frac{47 \pi ^2}{12}\right) \log ^2(\chi) \right.\nonumber\\
\left. +\left(-20 \zeta_3+6-23 i \pi -\frac{9 \pi ^2}{2}-\frac{26 i \pi ^3}{3}\right) \log (\chi) \right.\nonumber\\
\left. +\frac{28 \zeta_3}{3}+i \pi  (88 \zeta_3-65)+\frac{44 i \pi ^3}{9}-\frac{27 \pi ^2}{2}-\frac{1361 \pi ^4}{360}+18\right\}.
\end{gather}
The above result is then matched to the solution for the coefficient $c_{449,0,-1,0}$ in the point $s=-1$ and the limit $\chi=-t\rightarrow 0_+$, which determines the required integration constant.

We used the same method explained above to compute all ten constants that appear in branches of the non-planar master integral solutions with seven-propagators. Seven of these branches are of integrals with only one massive propagator and there the above method resulted in one-fold Mellin-Barnes integrals. For four of those seven branches (including the one explained above) we could close the corresponding contours and compute the final residue sums analytically with the XSummer package. The remaining three sets of one-fold Mellin-Barnes integrals were performed numerically with the package MBTools~\cite{MBTools} and we then used PSLQ~\cite{Ferguson91apolynomial} to match an integration constant onto a basis of transcendental constants up to weight four. Finally, the last three of the ten constants appear in branches of integrals with two massive propagators. These three constants were expressible in terms of two-fold and three-fold Mellin-Barnes integrals. We followed the steps explained in~\cite{Anastasiou:2013srw} to compute them, by mapping them onto parametric Euler-type integrals, at which point, they could be calculated using standard methods.

\section{Analytic continuation and crossings} 
\label{sec:crossings}

The top-sector master integrals that we include in our ancillary files are independent under crossings of the external momenta $p_{1,2,3}$ and may be directly used in the region where $p_{2,3}$ are incoming, i.e. $u>0$ and $\chi=t/s>0$. However, the physical two-loop virtual mixed QCD-EW amplitudes describing the process $i_2(p_2)i_3(p_3)\rightarrow i_1(-p_1)+V$ with coloured partons $i_{1,2,3}$, contain, after IBP reduction, also crossings of our chosen top-sector and lower-sector master integrals.\footnote{Furthermore, in order to compute all helicity amplitudes one requires also integrals where $p_{1,2}$ or $p_{1,3}$ are incoming, corresponding to $s>0,t>0$ respectively, for which one would need to analytically continue our master integrals.} As explained in Section~\ref{sec:masters}, all our master integrals are expressed in terms of GPLs with argument $\chi$. In order to see any possible cancellations among different crossed master integrals in the amplitudes, it is required to express them also in terms of GPLs with the same argument $\chi$. In this section we briefly explain how these crossings can be performed in practice. 

There are three physical scattering regions, defined in terms of the Mandelstam invariants
\begin{align}
(2a)_+ \; :& \quad s >0\,, \quad t,u < 0 \,,\\
(3a)_+ \; :& \quad t >0\,, \quad s,u < 0 \,,\\
(4a)_+ \; :& \quad u >0\,, \quad s,t < 0\,,
\end{align}
along with $m_V^2>0$. Since the master integrals are computed in Minkowski space, they are imaginary and their imaginary pieces are fixed by providing the corresponding positive Mandelstam variable with an infinitesimal positive imaginary part according to the Feynman prescription,
\begin{align}
(2a)_+ \; :& \quad s \to s + i\, 0 \,,\\
(3a)_+ \; :& \quad t \to t + i\, 0 \,,\\
(4a)_+ \; :& \quad u \to u + i\, 0 \,.
\end{align}
The integrals in the ancillary files included with this paper are all defined in the region $(4a)_+$. In Ref.~\cite{Anastasiou:2000mf} it is explained how to perform the analytic continuation from the region $(4a)_+$ to the other two regions $(2a)_+, (3a)_+$ and we refer to that paper for details.

In order to compute the physical scattering amplitudes in the region $(4a)_+$ (or the other two), one will need all possible permutations of the external momenta
\begin{eqnarray}
\sigma_{ij}&:=& p_i \longleftrightarrow p_j, \quad (ij)\in\{(12),(13),(23)\}, \label{eq:cross1}\\
\sigma_{ijk}&:=& p_i\rightarrow p_j, p_j\rightarrow p_k, p_k\rightarrow p_i, \quad (ijk)\in\{(123),(132)\}. \label{eq:cross2}
\end{eqnarray}
Consider now for example the crossing $\sigma_{12}$, which maps $\chi\rightarrow -1-\chi-\mu$ and the region $(4a)_+$ to the region $(3a)_+$. In order to compute the crossing $\sigma_{12}$ of our master integrals, we first map $\chi\rightarrow -1-\chi-\mu$ and expand the mapped  integrals in small $\mu$, resulting in GPLs with argument $-1-\chi$. Then we express these GPLs in terms of GPLs with argument $\chi$ using so-called {\it supershuffling} identities, as implemented for example in the Maple-based package HyperInt~\cite{Panzer:2014caa}. This supershuffling step needs to be done assuming one is in region $(3a)_+$ and results in all GPLs having the argument $\chi$. After this step, one follows Ref.~\cite{Anastasiou:2000mf} and analytically continues the final expression from region $(3a)_+$ back to $(4a)_+$ to get the final result for the $\sigma_{12}$-crossed master integrals as defined in the region $(4a)_+$. Similar steps can be performed for the other permutations in Eqs.~\eqref{eq:cross1},~\eqref{eq:cross2}.

We have implemented the above crossings and, in fact, used them to make various non-trivial checks of our master integral solutions gathered in the ancillary files, that we computed by solving the differential equations as explained in section~\ref{sec:masters}. Namely, many of the lower-sector integrals that we computed in the total list of $468$ master integrals are related by crossings of the external momenta $p_{1,2,3}$. By applying the above crossing steps, we have checked that these crossing relations are indeed satisfied by our master integral solutions.

\section{Validation of Master Integrals} 
\label{sec:checks}

We have performed multiple checks to make sure that the computed master integrals are correct. To begin with, two ``basic'' checks were performed for almost all master integrals. %
Firstly, the consistency of our analytic solution of a given master integral with its 
corresponding  $\mu$- and $\chi$- differential equations was checked. Secondly, upon deriving a boundary constant analytically with the methods of Section~\ref{sec:boundary_condition}, it was %
validated against numerical integration using the Mellin-Barnes method as implemented in MBtools \cite{MBTools}. In addition to these checks and the ones mentioned in the previous Section, %
we employ two other which are discussed below.

One straightforward and robust way of verifying our results is to compute the master integrals numerically and compare them with their analytic solutions at a kinematic point. %
To this end, we have used the computer programs pySecDec \cite{Carter:2010hi,Borowka:2012yc,Borowka:2015mxa,Borowka:2017idc} and FIESTA \cite{Smirnov:2015mct}, that are based on the sector-decomposition method, to compute integrals numerically. %
In this way, the master integrals with up to five denominators ($t=5$) were systematically compared against numerical results and the agreement at the per-mille level was obtained. However, it was already difficult to achieve this precision for some master integrals with five propagators, %
due to their specific kinematics. Typically, problematic integrals have threshold singularities that are hard to treat with the sector decomposition method. 

Even though there is an algorithm that is implemented in \texttt{PySecDec},\footnote{We thank S.P.~Jones and S.~Jahn for pointing out the possibility to handle threshold singularities within pySecDec and their help with our integrals. More details on threshold singularities can be found in Refs.~\cite{Borowka:2017idc,Borowka:2018goh}.} we opted for a different approach that is known in the literature~\cite{DiVita:2018nnh, Lee:2019lno} to deal with more sophisticated integrals ($t=6,7$). To demonstrate it, consider a Feynman integral $I^{d+2}(\Vec{x};d)$ in $d+2$ dimensions with $\Vec{x}$ being a vector of Mandelstam variables and $ d = 4 - 2\epsilon $. There exist dimensional recurrence relations (DRR) for dimensionally regularized Feynman integrals \cite{Tarasov:1996br} that read as
\begin{equation}\label{eq:checks_drr}
    I^{d+2}(\Vec{x};d) = \sum_{i} R(\Vec{x};d) I_i^{d}(\Vec{x};d),
\end{equation}
where $R(\Vec{x};d)$ are coefficients that are rational functions of Mandelstam variables that can be computed following Refs.~\cite{Tarasov:1996br,Lee:2009dh}; $I_i^{d}(\Vec{x};d)$ are Feynman integrals defined in $d$ dimensions which belong to the same integral \textit{family} as $I^{d+2}(\Vec{x};d)$.

\begin{figure}
    \centering
    \includegraphics[width=\textwidth]{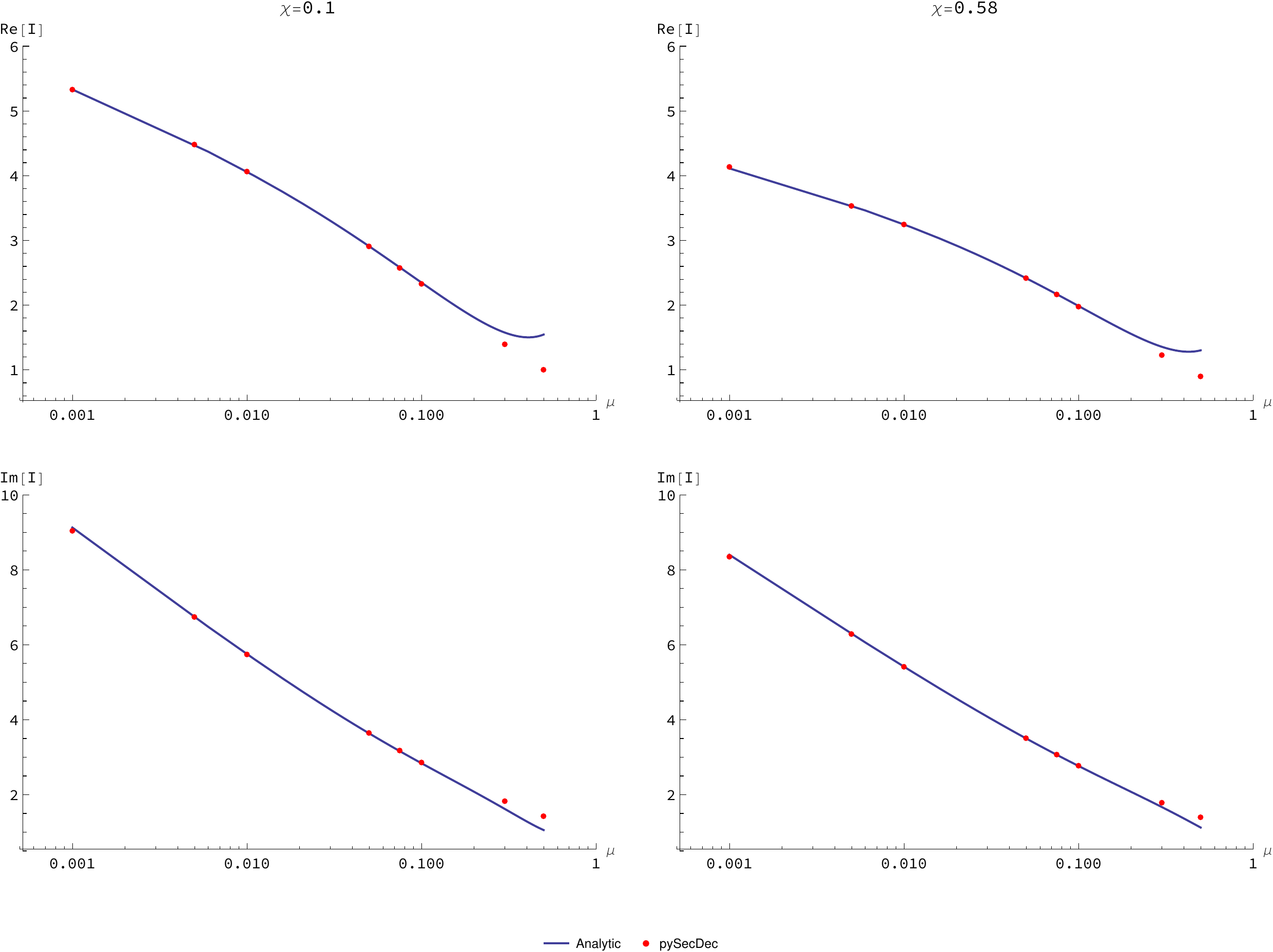}
    \caption{We compare the real (top row) and imaginary parts (bottom row) of the analytic solution (blue line) against the numerical results (red dots) for a finite integral $I^{\,\text{np2}}_{210110011}$ in $d = 6$ at different kinematic points. The numerical error bands are too small to be seen.}
    \label{fig:finite_check}
\end{figure}

If we choose $I^{d+2}$ such that it is \textit{finite} in the $\epsilon\to0$ limit, DRRs can be used to put constraints on lower-dimensional integrals. %
Indeed, if $I^{d+2}$ is finite, then at the right hand side of Eq.~\eqref{eq:checks_drr} all poles in $\epsilon$ have to cancel exactly. Hence, to use DRR constraints %
in our case, $I_i^{d}(\Vec{x};d)$ are expressed in terms of computed master integrals using integration-by-parts identities. The exact cancellation of poles on the right hand side of Eq.~\eqref{eq:checks_drr} %
provides a non-trivial check of the computed MI. To check finite parts of the master integrals, we computed finite $I^{d+2}$ using pySecDec. We note that DRRs do not resolve the problem with threshold singularities, but it was much easier to achieve per-mille numerical precision in the case of finite Feynman integrals.
An example of such a comparison is given in Fig.~\ref{fig:finite_check}. 

Various programs were used to pursue this check. To find $d=6$ and $8$ finite integrals, the reduction program Reduze2 \cite{vonManteuffel:2012np} that implements an algorithm of Ref.~\cite{vonManteuffel:2014qoa} was used. To generate the DRRs in Eq.~\eqref{eq:checks_drr},
we used the Mathematica-based package LiteRed \cite{Lee:2013mka}. To perform reduction to our master integrals, we employed both Kira \cite{Maierhoefer:2017hyi,Maierhofer:2018gpa} and Reduze2. %

\section{Conclusions} 
\label{sec:conc}

In this paper we computed the Feynman integrals needed for the double-virtual mixed QCD-EW NNLO contribution to $Z+j$ production at hadron colliders (disregarding contributions from the top and the Higgs, as discussed previously). The computation was performed as an expansion around $m_Z=0$, making the result valid in the high $p_T$ limit. This makes it useful for background determination for certain types of new physics, such as weakly interacting dark matter (see ref.~\cite{Lindert:2017olm} and the references therein).

Our results are presented as an expansion in $\mu=-m_Z^2/s$ of the form given by eq.~\eqref{eq:muansatz}, keeping terms up to $\mathcal{O}(\mu)$, and with the coefficients of the expansion being expressed in terms of generalized polylogarithms. The results are added as an ancillary file as described in Section~\ref{sec:anc}. We obtain numerical agreement with SecDec (with the help of finite basis methods) as described in Section~\ref{sec:checks}. If higher accuracy is needed, there is no conceptual barrier to extending the $\mu$-expansion to higher orders.

Additionally we have computed the integrals (when normalized as in eq.~\eqref{eq:idef}) up to terms of transcendental weight $4$. This is expected to correspond to the terms contributing to the amplitude up to finite orders in $\epsilon$. It could, however, happen that intricate cancellations take place in such a way as to make higher $\epsilon$-orders for some individual integrals needed. In that case there is nothing conceptual that prevents us from continuing the expansion to the required $\epsilon$-order.

An obvious next step, is to use the integrals computed in this paper, to perform the complete calculation of the NNLO mixed QCD-EW correction to the double-virtual scattering amplitude for $Z+j$ production. Additionally, one has to compute the real-virtual and the double-real contributions, before being able to perform the infrared subtractions and combine it all into a final result for the scattering cross-section. Moreover, we may be able to estimate or even compute the top-mass contribution that were disregarded in this paper as discussed in the introduction. Finally, we note that the integrals computed in this paper have a large overlap with those needed for $W_{\pm}+j$ production, with only a few extra integral families missing. All of this may be the subjects of future publications.

\acknowledgments
We are very grateful to Kirill Melnikov for the idea and inspiration to pursue this project, as well as for numerous enlightening
discussions and for many helpful comments on the manuscript.
We thank S.P.~Jones and S.~Jahn for their help with pySecDec \cite{Borowka:2017idc} and J. Usovitsch for his help with Kira \cite{Maierhoefer:2017hyi}.  
We also thank V.A. Smirnov for help with the results of Ref.~\cite{Henn:2015sem}. The research of C.W. was supported in part by the BMBF project No. 05H18WOCA1.
The work of H.F. is part of the HiProLoop project funded by the European Union’s Horizon 2020 research and innovation programme under the Marie Sk{\l}odowska-Curie grant agreement 74717. The  research  of  K.K.  was  supported  by  the  DFG-funded  Doctoral  School KSETA (Karlsruhe School of Elementary Particle and Astroparticle Physics) and by the European Research Council under the European Unions Horizon 2020 research and innovation programme (grant agreement 740006).

\appendix
\crefalias{section}{appsec}
\bibliographystyle{apsrev4-1} %
\bibliography{main.bib}

\end{document}